# Building ventilation : a pressure airflow model computer generation and elements of validation

H. BOYER, A.P. LAURET, L. ADELARD, T. A. MARA

*Université de la Réunion, Laboratoire de Génie Industriel, Equipe Génie Civil, BP 7151,*

*15 avenue René Cassin, 97705 Saint-Denis Cedex, Ile de la Réunion, FRANCE*

Abstract :

The calculation of airflows is of great importance for detailed building thermal simulation computer codes, these airflows most frequently constituting an important thermal coupling between the building and the outside on one hand, and the different thermal zones on the other. The driving effects of air movement, which are the wind and the thermal buoyancy, are briefly outlined and we look closely at their coupling in the case of buildings, by exploring the difficulties associated with large openings. Some numerical problems tied to the resolving of the non-linear system established are also covered. Part of a detailed simulation software (*CODYRUN*), the numerical implementation of this airflow model is explained, insisting on data organization and processing allowing the calculation of the airflows. Comparisons are then made between the model results and in one hand analytical expressions and in another and experimental measurements in case of a collective dwelling.

# 1- Introduction

Regarding the number of airflow network models found in building publications, it appears that the calculation of airflow is of great importance for detailed building airflow network modelling computer codes. In fact, these airflows most often constitute an important thermal coupling between the building and the outside and the different thermal zones. It is thus, that during the period when heating a residential building, approximately 30% of the energy loss is due to air renewal[1]. Furthermore, the airflow exchanged between two different zones, separated by a standard doorway with a temperature difference of 0.1 K is approximately 120 m$^3$/h [2]. These airflows also play a major role in humidity displacement in the building (comfort, localised condensation risks, ...), the dispersion of pollutants [3,4] (smoke, gas flows ...), or in industrial [5] and farming buildings[6].

The content of this paper is macroscopic in that it does not determine the temperatures, the pressures, or the average speeds for all points in the zone. The above approach can be arrived at by using simplifications associated with the Navier-Stokes infinitesimal volume equations [7]. In our case, the aim is to obtain an estimation of the exchanged airflows.

Depending on the climate, these airflows must be considered differently. Thus in tropical climates, natural ventilation affects essentially the inside comfort by favouring evapo-perspiration through the air movement. The buildings in these latitudes are often very much open to the exterior and thus the airflow transfer aspect is extremely important. In temperate and cold climates, the energetic aspect is the determining factor in these airflow transfers. In fact the importance of ventilation in the energy consumption continues to rise due to better building insulation.

## 2 – The driving effects

As these notions have already been largely studied [9,10], only the necessary explanations for the comprehension of the paper will be covered. As well as wind and thermal buoyancy, mechanical ventilation by air extractors constitutes a motor of which airflow and pressure characteristics are taken as known.

2.1 The wind :

Due to the nature of the data available and the calculation time necessary, the model does not take into account the fluctuations of these parameters. In fact two problems arise : firstly for the considered site which has no experimental data, the wind characteristics are those registered by the nearest weather station. Secondly, the wind data available is generally only that which is recorded and averaged over a certain period of time (most frequently over one hour). The evolution of airflows in a building is thus assimilated to a succession of steady states. Each of these is considered to be a state in which the parameters (wind speed and direction, temperatures and mechanical ventilation airflows) are averaged over the considered time period. Other approaches are being investigated [11]. With the above restrictions the wind exerts on an obstacle, at the height z, an aerodynamic pressure given by the following equation:

$$\overline{P(z)} = \frac{1}{2} \rho \; \overline{Cp} \; \overline{v}(z)^2$$

The average pressure coefficient, dependent on wind direction, represents an exerted pressure if the coefficient is positive and a depression when negative. These coefficients stem from ventilation fan experimental studies [12, 13] or numerical simulations.

2.2 Thermal buoyancy :

In the case of a still fluid, such as air in a room (without atmospheric disturbance), there exists a pressure gradient created only by a unit weight field so

$$\frac{dP}{dz} = -\rho g$$

Moreover, the air unit weight is related to the temperature through the following equation :

$$\rho(T) = \rho(T_\circ) \frac{T_\circ}{T_\circ + T}$$

Thus, an airflow will be created through an opening separating two air volumes at different temperatures, due to the pressure differences from one part to another.

### 3 – The effects coupling :

In a building, at each moment, the two previous effects combine to create an airflow distribution. A detailed study of effects superposition is proposed by Sherman [14]. In certain individual cases, some aspects of this combination are predictable. It is thus that thermal buoyancy is most important in a high building or will be dominant over wind speed when this last factor is low. The analytic study can only be carried out in simple cases, concerning a monozone building with a small number of openings [15]. The reference pressure of the room being unknown, the solving of the air weight balance gives the internal pressure and therefore the airflow. This analytic approach reaches its limits when taking into account mechanical ventilation [16], and therefore leaves only numerical solutions for multizone buildings.

3.1 The setting up of the airflow pressure model :

In reality, the thermal and airflow phenomena are related. The solving of this coupling can be achieved by establishing a system (called coupled) which regroups, on the one side the equations related to the thermal system (determining the dry bulb temperature of the different thermal zones), and on the other, the overall weight balance which enables the determination of the flows. The arrival at a coupled mathematical model, if at all possible [16, 17], is not compatible with our objective to set up a controlled resolution design tool [18]. On the other hand, our prior choice to solve the thermal coupling of zones through an iterative schema [19], imposes the choice of an iterative coupling between the thermal and airflow phenomena. In fact, we want to draw up from this iterative schema the possibility of adjusting the accuracy of the model (through the possibility of choosing models more or less precise [20]).

The following paragraph illustrates a multizone building, comprising only small openings. Through this type of opening, once the pressures are established, the air circulates in only one direction. A relation exists between the airflow and the pressure differences, known the Crack Flow equation.

$$\dot{Q} = Cd\ S\ (\Delta P)^n$$

The problem therefore exists in the determination of the discharge coefficient (*Cd*) and the exponent (*n*). For this last point, when the viscosity forces are dominant the flow rate can be considered as laminar, and the value to choose for n is 1. When the dynamic forces are predominant, the flow rate is turbulent and thus, n values 0.5. The average value of 0.67 is considered as valid for the different pressures found in housing with classic small openings. The airflow rate coefficient, experimentally measured, varies considerably with the geometry of the opening. To generalise, the product of this coefficient and the surface of the opening, has a permeability dimension. The previous equation thus becomes $\dot{Q} = K\ (\Delta P)^n$, where K represents the mass flow through an opening for a pressure difference of 1 Pa.

Owing to the thermal and airflow decoupling, mentioned previously, the dry air temperatures of each zone are taken to be known. The unknown factors to the problem are therefore the reference pressures for each one of the zones, this reference pressure being for example at floor level. From henceforward in our paper, a small opening which leads to the outside shall be called exterior. In the case of an interoom opening it shall be called interior. The pressure difference relations are obtained by the application of the Bernoulli equation on a flow line, for an incompressible and non viscous fluid (the variation of the air speed from one side to the other of the opening is supposed to be neglectable).

$$\frac{1}{2}\rho v^2 + P + \rho g z = Cste$$

Therefore, the following schemata and equations can be applied for the two previous cases.

Fig 1 : *Exterior and interior small openings*

$$\Delta P = Pae + \frac{1}{2}\rho(T_{ae}) C_p v^2 - \rho(T_{ae}) g z - (Pz(i) - \rho(T_{ai(i)}) g z)$$

and

$$\Delta P = Pz(j) - \rho(T_{ai(j)}) g z - (Pz(i) - \rho(T_{ai(k)}) g z)$$

Taking into account the inside air volume incompressibility hypothesis, the mass weight balance should be nil. Thus, in the established system, called pressure system, the unknowns are the reference pressure for each of the zones. Mechanical ventilation is simply taken into account through its known airflow values. In this way, ventilation inlets are not taken into detailed consideration which could have been the case owing to their flow-pressure characteristics [21]. Considering $\dot{Q}_{i,j}$, the airflow rate from zone i to zone j (suffix 0 representing the outside) the system is written as follows

$$\begin{cases} \sum_{i=0,i\neq 1}^{i=N} \dot{Q}(i,1)+\dot{Q}_{vmc}(1)=0 \\ \sum_{i=0,i\neq 2}^{i=N} \dot{Q}(i,2)+\dot{Q}_{vmc}(2)=0 \\ ... \\ \sum_{i=0}^{i=N-1} \dot{Q}(i,nZ)+\dot{Q}_{vmc}(N)=0 \end{cases}$$

The airflows depending on pressure differences through the fractional exponent, it constitutes a non-linear system and therefore its calculation requires the use of the appropriate solving techniques. This point is covered after the representation of the airflow network and the description for the case of large openings.

3.2 The nodal pressure representation :

The previous equations enable the setting up of an analogic network, representing the problem. Each of the reference pressures of the zone as well as the outside pressure correspond to a node in the network. Conductances linked to the wind or to thermal buoyancy are placed between the pressions For the simple building (taken to comprise only small openings) in the following figure, the corresponding analogic network is associated :

*Fig 2 :Airflow network*

The hypothesis underlying this network model, composed only by conductances, is that of a steady state. In fact, for each time step, the pressure field considered is obtained by the supposition of sollicitations being constant over this period of time. In a dynamic case, each pressure node is associated with an airflow capacity, which, in the electric schema, leads to taking capacities into account. However, taking into account the value of the time constants inserted ($10^{-1}$ to $10^{-2}$ s), before those of the parameters (a few tenths of a second for the wind) the *transitory state* becomes practically non-existent and the system finds itself almost instantly in steady state. [2]. As regards the pressures, the building will react following the outside sollicitations. The consequences which arise and affect the solution are covered in the chapter linked to the numerical aspects. This airflow model nodal approach enables the setting up of the model with the help of matricial algebra and to build up, in a systematic way, the system to be solved [22], a method which we held on to, whilst working out our code.

3.3 Large openings taking into account :

The integration of large openings (doors, windows, ...), represents a problem in a calculation code, because of the possibility of simultaneous incoming and outgoing flows due to the driving effects superpositions. The airflow equation is only valid in the only case of wind effect, in which case the discharge coefficient is given, for example for bay windows in [23].

For a multizone building, an approach is outlined by Ansley [24]. For a long time this particular point has been a stumbling block in multizone models. The first study dedicated to the mass transfer through large openings was that of Brown et al [25, 26]. Besides certain analytical approaches [27, 28], the most frequently used approach is that of non-dimensional experimental correlation [29, 30]. Once established, there arises the problem of the coupling of the large openings to an airflow calculation code for which we present the basics.

With the calculation for small openings being simple, the idea is to divide the large opening into various small openings, for which the airflow equation can be applied. Walton [31] shows that the division of a large interior opening into two small openings placed respectively at a height of *5/18* ths.and *13/18* ths, leads to a relation similar to the correlation shown previously ($Nu/Pr = \frac{C}{3}\sqrt{Gr}$, *C* coefficient depending on the involved characteristics temperature difference) in the condition that a discharge coefficient of *0.78*, and an airflow exponent of *0.5* are used for each of the small openings. The same approach is adopted for external openings, but leads to the problem of coupling with the wind. This is realized through pressures, but nessitate to use wind link pressure at level of the large opening, what is physically problematic. Meanwhile, this approach is the one choosen by many authors [32, 33, 34]. The tests we made shows taht it is acceptable when airflows through these external openings are in one way only and if the permeability of the concerned walls is too important (30% being a practical limit).

The advantage, which leads us to integrate the large opening considered as two individual small openings (layout, surface, discharge coefficient and exponent) into this model, is the possibility to link directly the large opening to the airflow pressure system outlined in the previous paragraph. Despite the practical aspect of this approach, it is never the less the approach of a pressure model and this sometimes raises problems due to the small pressure differences found. In certain cases, especially with large interior openings, the problem of the convergence speed arises, thus necessitating further particular techniques for numerical resolution.

## 4 – Considerations on the solution of non linear pressure system :

With the selected models for large and small openings, the writing of the mass balances of each of the zones leads us to a non-linear system, where the unknowns are the reference pressures of the zones, constituting a vector **p** = $(P_1, P_2, ...)$. For a zone suffixed i, the mass balance written $f_i(\mathbf{p})$ should be nil. For the whole building, $\mathbf{f}(\mathbf{p}) = 0$ must be solved when writing the system in vectorial form. A review of airflow simulation codes [35] shows that the method most frequently used for solving this system is that of Newton-Raphson, employed mostly with the addition of a few modifications [33]. This method, shown in detail in ref [36], illustrates that the previous system gives the following equation :

$$\mathbf{p}^{n+1} = \mathbf{p}^n - \frac{f(\mathbf{p}^n)}{J(\mathbf{p}^n)}$$

The Jacobian matrix is formed with the zones balances partial derivatives regarding the reference pressures. Thus for example, for a building comprising four zones the matrix is written :

$$J(\mathbf{p}_n) = \begin{bmatrix} \frac{\partial f_1}{\partial p_1} & \frac{\partial f_1}{\partial p_2} & \frac{\partial f_1}{\partial p_3} & \frac{\partial f_1}{\partial p_4} \\ \frac{\partial f_2}{\partial p_1} & \frac{\partial f_2}{\partial p_2} & \frac{\partial f_2}{\partial p_3} & \frac{\partial f_2}{\partial p_4} \\ \frac{\partial f_3}{\partial p_1} & \frac{\partial f_3}{\partial p_2} & \frac{\partial f_3}{\partial p_3} & \frac{\partial f_3}{\partial p_4} \\ \frac{\partial f_4}{\partial p_1} & \frac{\partial f_4}{\partial p_2} & \frac{\partial f_4}{\partial p_3} & \frac{\partial f_4}{\partial p_4} \end{bmatrix}$$

The previous equation can also be written $J(\mathbf{p}^n)\mathbf{D}^n = -\mathbf{f}(\mathbf{p}^n)$

$\mathbf{D}^n = (\mathbf{p}^{n+1} - \mathbf{p}^n)$ is the corrective terms vector to be multiplied to the previous pressures over the given time period to estimate the researched pressures. A consequence of the truncation of term with order more than one in the developement $\mathbf{f}(\mathbf{p}^n)$ to obtain the starting equation is that the pressures found when solving the linear system are not the final solutions to our problem. These values are only an approximation and therefore an iterative method is needed to reach the desired solution. The use of this method requires further explanation for its application to our pressure system

Thus the method leads only to quadratical convergence when the estimate is close to the solution. In our case, the physical analysis of the problem leads us to consider the evolution of the pressures as a succession of permanent states. In the majority of publications relative to airflow systems solving methods the previous time lapse pressure vector is used for initialising the iterative procedure. In the case of important pressure variations between two different time periods (due to the wind or imposed airflows from mechanical ventilation ), it can arise that the previous time lapse pressure vector is outside the convergence field. Walton therefore puts forward a method of pressure vector initialisation by linearising all the airflow equations (the airflow exponent is taken as equal to 1) the initial pressure vector considered is the solution to a linear system which characterises the laminar state in the building (but taking into account the wind effects and the thermal buoyancy) [37]. Furthermore, the integration of large openings into a pressure system can cause problems of convergence speed. Located between a zone and the outside, a large opening links greatly, the inside and the wind pressures. As in between two time lapses, the wind speed and direction can change considerably, the remarks of the previous paragraph apply. Another source of problems is the value of 0.5 given to the airflow exponent for the small openings equivalent to Walton's model. If the mass balance is symmetrical in relation to the pressures, and that the exponents are equal to 0.5 the Newton method diverges irremediably [38]. Feustel insists that the convergence of the method lowers as the number of exponents equal to 0.5 grows [35].

These divergence problems being ignored for the moment, the large openings may also compromise the speed of the convergence. It is furthermore established that a small pressure difference generates important mass flows, through a large opening. For the considered zones the mass balances partial derivatives have important numerical values (compared to a case concerning only small openings). Consequently, the amplitude of the successive corrective terms is low. In these conditions, an important number of iterations are necessary to reach the solution. As well, for a building which comprises various zones, separated by large openings, there thus exists various directions in which the convergence is slow. Between two time lapses, the distance between the pressure vectors is a function of the disturbance gradient due to the solicitations. The number of iterations can therefore be very important and also change considerably from one time period to another (this will be illustrated further).

## 5. Computer model implementation :

The former airflow model is a part of a detailed combined building thermal and airflow simulation code called *CODYRUN*, usable for building design. Previous articles have already analysed and retained the multiple model approach [39] and thermal implementation [19], so here, we will focus our attention upon the airflow module. Thermal and airflow modules are coupled each other thanks to an iterative proceed, the coupling variables being the air mass flows.

Fig. 3 : *Coupling flow chart*

It is useful to code the interzones flows with the help of a matrix $\dot{Q}$. The terms $\dot{Q}[i, j]$ then qualifies the mass transfer from the zone *i* to the zone *j*, as it is depicted at fig. 4.

Fig. 4 : *Interzonal airflow rates matrix*

Another way to code the mass transfer is proposed in [10], as interesting interpretation of the inverse matrix (age of air, …).

### 5.1 Associated data structures :

CODYRUN code being used in cases of aided design (non building specific code), it is then necessary, considering the description of any building, to establish the airflow model to determine the mass airflows. In a classical way, this description needs a number of zones, and of course, a number of aeraulic components, little small openings (called PO), or large openings (called GO), mechanical ventilation inlets, or imposed flows (called VMC). The description of the buildings is made with help of Microsoft Windows environment, as detailed in [40]. In a schematic way (fig. 5), the description generates description data files (building, zones, walls, windows, openings, ….) or structures, in which we will find all information fields linked to theses datas.

Particularly, the structure called *Opening* will contain, for each PO, a line constituted firstly with zone numbers between which this opening is set (for instance, *z1* and *z2* , number 0 is kept for the exterior), secondly with the quote z relative to the reference height to which the different zones pressures will be calculated, and then, parameters of the physical law associated to air flows, i.e. for instance *k* and *n*, respectively the permeability and the exponent of the flow). For a GO, we will find the zone numbers, height *h*, width *w* of this opening, quote, as well as the discharge coefficient *Cd* (in case of other model than Walton's).

For a VMC, the number of the concerning zone (from which the air is extracted) and the flow value (with the simplified hypothesis that this flow doesn't depend upon pressures from each part of the opening) are needed.

Fig. 5 : *The description windows, files and openings data strucrure.*

Considering Walton's model, chosen for modelling flows in the case of wide openings, each line is blasted in two (a line for each small opening), with parameters (quote and coefficient) corresponding to this model). Then, for each of these lines, are added an incremented number (whose purpose is to index the structure) and a number coding the type (PO, GO, or VMC) allowing to accelerate the further treatments.

## 5.2 General calculation flow chart :

Now, we are going to describe the treatments allowing to obtain the flow rates. They respond to the tasks represented at fig. 6. To each time step, the first modules leads to the calculation of the exterior pressures for each opening with the exterior, taking the wind (speed and direction) and the dry exterior air temperature into account. For a systematic calculation, an array of external pressures is used, indexed with the incremental number of openings. The following module leads to the determination of interior reference pressures and will be presented in the next paragraph. With help of the final pressures, the elementary flow rates through each of these openings are determined and the matrix of interzonal airflow rates is then filled.

Fig. 6 : *Flow chart for airflow rates calculation*

## 5.3 Detail of pressures calculation :

The non linear system being solved with an iterative scheme, the calculation of pressure for each zone needs the use of many matrix or vectors. The objective is to calculate $p_n$, vector of reference pressures at the present time step. $p_{n-1}$ is then the vectors including reference pressure at the previous time step. The number of iterations being represented by *k*. so, $p_n^k$ is the vector pressure of the iteration *k*. The most simple stopping criterion, relative to the pressure convergence of a pressure based norm (quadratic for example) is less interesting than this based on the exactitude of the mass flow balance for each zone. It is this last criterion we used in our study.

Fig. 7 : *Non linear pressure system solution*

The calculation of the vector of corrective terms needs at each iteration the calculation of the jacobian *J* and the vector of residuals balances *B*. The resolution of the pressure system needs the calculation of partial derivatives $D^k$ of the balance sheet for the filling of the Jacobian. The method chosen here use data structures previously described and the knowledge of the flow analytical expression and its derivatives. Actually, the mass balance for a zone is considered as a sum of terms similar to $k \Delta P^n$ and of constant terms linked to the VMC. Consequently, partial derivatives can be calculated. For instance, to calculate the term*[i, j]* from the jacobian matrix (derivative of the mass balance of the zone *i* toward the reference pressure of the zone *j* ), a reading of the *Opening* structure leads to know if there is an opening between the zone *i* and *j* (by looking at *z1* and *z2* fields, fig. 5) If there is one, we can write :

$$\frac{\partial f_i}{\partial p_j} = \frac{\partial}{\partial p_j}\left( \sum k(p_j - p_n)^n + \dot{Q}_{vmc} \right)$$
$$= -k n (\Delta P)^{n-1} = -n \frac{k \Delta P^n}{|\Delta P|}$$

If there are several openings, a sum must be made. The last equality lets appear at the numerator a term evaluated in the previous iteration and permits to replace a raise to an exponent (which consumes much calculation time) by a simpler division. For a very little pressure difference, the derivative is not defined. We linearize the flow expression ( *n=1* ) and the term of the jacobian turns to be equivalent to *–k*.

Doing the same for the filling of residuals balance, the organisation chart for the calculation of pressure is the following.

Fig. 8 : *Pressure calculation flow chart*

The resolution of the linear system use LU decomposition [41]. However, it has been necessary to transform the proposed source code in regard to double precision float coding.

## 6. Elements of airflow model validation

The experimental validation for the code has been conducted in laboratory conditions and in natural environnment, mainly for the thermal aspects.

For the airflow aspect, confrontations between the results of the code, and analytical expressions (for simple cases), expermenal values and other models or other airflow codes, have been made. Only the two first approaches are illustated in this paragraph.

6.1 Analytical checking :

In simple cases, using only small openings and mechanical ventilation, analytical expressions can be found [27][10]. All the cases used are available in [44] and we will illustrate here the combination of wind effects and mechanical ventilation.

For the simulation, we will study for a given enclosure, several different configuration (in steady state), by combining different weather sequences.

The artificial meteorological files (hourly datas in that case) used for the simulation is constituted with the following characteristics: Global and diffuse radiation are null and outdoor dry air temperature constant and equal to 25 °C. A wind with a southern direction (speed of 2m/s) is maintained in all the simulation day.

For the enclosure, indoor dry air temperature is maintained at 0°C and two small openings are locared on the southern wall.

. The first one is at $z_1$=0.5m and the second is at $z_2$=2.5 m. For each of these openings, parameters relative to the flow are permeability and (K=0.5) and the exponent of the flow (*n*=0.67). In this case, the analytical expression for the flow rate is written :

$$\dot{m} = \rho_{ae} K \left( -\left( \frac{\rho_{ae} - \rho_{ai}}{2} \right) g(z_1 - z_2) \right)^n$$

The rate is calculated with the previous expressions is equal to *0.497* kg/s and is practically equal to that given by the model after convergence.

To make a second test, one of the two little openings of the previous building is disposed on the northern wall. In order to consider only the effect of wind and mechanic ventilation, the indoor air dry temperature is maintained to the same value tha outdoor, i. e. 25°C. To test the influence of the mechanical ventilation, we use a mechanical ventilation (VMC) extracting 3000 m3/h.

To study different working rate, we have associated a hourly profiles for extacted flow :

Null before 7 hour in the morning (to test only the wind influence), equal to 10% of the nominal value before 13 hours, to 100% until 18 hours and null for the rest of the day. We represent at the figure 9 the different calculated mass flows for the building (the positive sign correspond to outgoing flows).

Fig. 9 : *Evolution of outcoming airflows*

The previous graph makes appear the three different cases :

- 0∏6 hour and 19∏24 h : When there is no mechanical ventilation (and no thermal bouyancy because of the temperature set value), the flow through small openings are due to wind. Air is getting into the room by the exposed opening.

- 7∏12 hour : In this case, the extraction rate (0.108 kg/s) is low compared to the previous rates. The mechanical ventilation is not greatly affecting the indoor pressure. The air extraction creates a depression in the room, that tends to raise the entering flow, so to decrease the outgoing one.

- 13∏18 h : Due to the importance of the extracted flow, the mechanical ventilation imposes the internal pressure. The depression is such that an air inlet appears on the opening not exposed to wind.

As an illustration of the numerical process, we show at figure 10 for the same day of simulation the number of iteration needed for the resolution of the airflow system. The number of iteration depends of course on the chosen convergence criterion, but the shape of the curve remains the same.

Fig. 10 : *Required iterations*

To each induced perturbation caused by the mechanical ventilation (to 7, 13, and 18 h.), the interior pressure evoluates and numerous iterations are needed. For the other hours, solicitations being constant (due to the constitution of the meteorological file), it is not necessary to iterate.

6.2 Experimental confrontation :

Due to the difficulty of airflow instrumentation for a building with wide openings, we have experimented from the thermal aspect a typical dwelling of collective building in Reunion Island. In this paragraph, we won't present the whole experimentation. The objective is only to show the coherency of results. Considering the coupling between thermal and airflow aspects (particularly for windy conditions, and for a building highly radiatively sollicited), during simulation, a poor determination of the mass air flow rates will lead to errors on indoor temperature and relative humidity. This instrumentation is part of technical evaluation of the building prescriptions ( ECODOM[45] ) for French overseas territories. The dwelling, represented at figure 11 includes three bedrooms and a living room. It is situated under roof.

Fig. 11: *Typical dwelling*

Different measurement sequences have been organized. First, measures with the dwelling totally closed have been made (doors and windows sealed), leading to thermal model validation independently of the airflow aspect. Then, different airflow scenario has been realised. This operation has been facilitated by the sliding nature of openings (glass windows or plain doors ). We will expose one of these scenario in this paper. The dwelling will be considered to be composed of five thermal zones (three bedrooms, living room, kitchen, bathroom + lavatory). In each of these zones, dry air and resultant temperature (in different points), relative humidity, some surface temperatures (inside and outside the dwelling for ceiling and walls exposed to solar radiation), as well as global radiation for the West side has been measured.

For the meteorological data, a station has been placed on the roof of the dwelling, giving semi-hourly datas (each 30 mn) for outdoor dry air temperature, sky temperature, relative humidity, global and diffuse horizontal radiation and also wind speed and direction.

For the presented sequence (9 days), all the doors and windows has been closed, excepted :

- Inside the dwelling, the open sliding bay window between berdoom 2 and the living-room (considered as a large indoor opening).
- Outside the dwelling, squared openings of 12 cm in the eastern wall of bedroom 2 and the western wall of the living-room (forming two small external small openings).

Considering mechanical ventilation in certain rooms (bathroom, lavatory, kitchen) and air leakage for the openings considered as closed, the non linear system which composes the pressure system has the same dimension as the number of zones, five. On the whole sequence, the following figure shows the flow rates (in $m^3/h$) coming through the living room from the outdoor environment via the western small opening.

Fig. 12 : *Incoming air flow in the living room*

Finally, comparisons of dry temperatures for one of the room is proposed at the figure 13.

Fig .13 : *Dry air temperature drawings*

Dry air temperatures has been measured in two points of the room, by means of thermocouples protected from radiation. However, because of the East-West orientation of the dwelling, solar beam radiation is entering the living room (via the bay windows) and so affects the measures (specailly days 3, 6 et 7), explaining the measured temperatures peaks (about 17 h) that are not described by simulation. These same observations have been made while the experimental sequence where the dwelling was totally closed. So, considering the sensor's precision, a good agrement between the model and the measures is clearly reached. On other sequences, in particular with large external openings, (and so with more important airflow rates) the results of simulation are still good and compatibles with our main objective of design tool.

## Conclusion

Because of the necessary coherency between different levels of model, the taking account of multizone characteristic in buildings is combined with a detailed calculation of airflow rates. We exposed different approaches and explicated the reasons that lead us to the choice of models used here. So, a similar treatment of wide and little openings is really possible, but in certain cases, numerical difficulties can appear. Some of these have been evocated in this paper. However, it is important to point out the great progress brought by the realistic taking account of the airflow rates in building simulation codes. The airflow model presented here is a part of the software *CODYRUN*. In particular, this application developed as a modular and evolutive structure will allow us to include further model studies actually conducted for large openings [33].

REFERENCES :

1. Pelletret, R., Khodr, H. *Transferts d'air entre pièces*. Revue Générale de Thermique, 1989, n°335-336. p. 657-662.

2. Passard, J., Peube, J.-L. *Modélisation des phénomènes érauliques dans l'habitat et méthode de réduction*. Revue de Physique Appliquée, 1990, n°25. p. 81-98.

3. Waters J.R., Simons M.W. *The evaluation of contaminant concentrations and airflows in a multizone model of a building*, Building and Environment 22, 4, 305-316 (1987)

4. Lansari & al, Dispersion and automotive alternative fuel vapors within a residence and its attached garage, Indoor Air, 6, 118-126

5. Bring, T.G. Malmström, A. Boman, Simulation and mesurement of road tunnel ventilation, *Tunelling and Underground Space Technology*, Vol. 12, N° 3, pp. 417-424.

6. J.M. Bruce, natural convection through openings and its application to cattle building ventilation, J. Agr. Eng. Res., 23, pp.151-167 (1978)

7. T.E. Kreichelt, Natural ventilation in hot process buildings in the steel industry, Iron steel Engineer, Dec 76, pp. 39-40.

8. Marenne, C., Fragnaud, F., Klammer, J., Groleau, D *Modélisation de la ventilation naturelle à l'intérieur des locaux en climat tropical : couplage thermique-ventilation*. France : CERMA, 1985, 53 p. Rapport n° 5.04.1010.

9. M.D. Lyberg, Basic air Infiltration, *Building and Environment*, Vol. 32, N° 1, 1997

10. D. Etheridge, M. Sandberg : *Building ventilation; Theory and measurement*, ISBN 047196087X, John WILEY & Sons, England, 1996

11. Haghighat F., Rao J., Fazio P. The influence of turbulent wind on air change rates - a modelling approach. *Building and Environment* 26, 2, p. 95-109 (1991)

12. Vickery B.J., Karakatsanis C. *External wind pressure distributions and induced external ventilation flow in low-rise industrial and domestic structures*. ASHRAE Transactions, 1987, Vol. 93. p. 2198-2213.

13. Gandemer J. Champ de pression moyenne sur les constructions usuelles. Application à la conception des installations de ventilation. Cahiers Scientifiques et Techniques du Bâtiment, N° 1997, 35 p (1987)

14. Sherman, Superposition in infiltration modelling, *Indoor Air*, 2, pp. 101-114, 1992.

15. Cadiergues, R. *Méthodes d'étude de la ventilation naturelle.* Promoclim E, Etudes Thermiques et Aérauliques, 1977, Tome 8E n°5. p. 307-318.

16. Allard, F., Herrlin, M. *Wind-Induced Ventilation.* ASHRAE Transactions, 1989, Vol. 95, Part. 2. p. 722-728.

17. Axley, J. *The coupled airflow and thermal analysis problem in building airflow simulation simulation*. ASHRAE Transactions, 1989, Vol. 95, Part. 2. p. 621-628.

18. P. Schmidt Schneider, J.J. Roux, J. Brau, Strategies for solving the airflow thermal problem in multiroom buildings, *Building and Environment* **30**, 2, p. 277-286 (1995)

19. Boyer H, Chabriat J.P., Grondin-Perez B., Tourrand C., Brau J. Thermal building simulation and computer generation of nodal models, *Building and Environment, 31, 3, pp. 207-214, (1996)*

20. H. Boyer, Brau J., Gatina J.C., Multiple model software for airflow and thermal building simulation. A case study under tropical humid climate, in Réunion Island. *In Proceedings of Building Simulation' 93*, IBPSA, Adelaide, Australia (1993)

21. Walton, G.N. *Airflow network models for element-based building airflow modeling.* ASHRAE Transactions, 1989, Vol. 95, Part. 2. p. 611-620.

22. Haghighat F., Rao J. Computer aided building ventilation system design - a system-theoretic approach. Energy and Buildings, 1991, n°17. p.147-155.

23. Provan T.F., Younger J.D. Air infiltration characteristics of windows. *Energy and Buildings* 9, p. 281-292 (1986)

24. Aynsley, R. M. *A resistance approach to estimating airflow through buildings with large openings due to wind.* ASHRAE Transactions, 1988, Vol. 94. p. 1661-1669.

25. Brown, W.G., Solvason K.R. *Natural convection through rectangular openings in partitions 1 - Vertical partitions*, Int. J. Heat and Mass Transfer, 1962, Vol. 5. p. 869-881.

26. Brown, W.G. *Natural convection through rectangular openings in partitions 2 - Horizontal partitions*, Int. J. Heat and Mass Transfer, 1962, Vol. 5. p. 869-881.

27. Allard, F., Utsumi, Y. *Airflow through large openings*. Energy and Buildings, 1992, n°18. p.133-145.

28. Van Der Maas, J. *Air flow through large openings buildings.* Lausanne: International Energy Agency Annex 20 substask 2 technical report, LESO-PB, EPFL, 1992. 163 p.

29. Khodr-Mneimne, H. *Transferts thermo-aérauliques entre pièces à travers les grandes ouvertures.* Thèse : Sci. : Université de Nice, 1990. 120 p.

30. M. Santamaouris, A. Argiriou, D. Asimakopoulos, N. Klitsikas, A. Dounis, Heat and mass transfer through large openings by natural convection, *Energy and Buildings* **23**, 1-8 (1995)

*31.* Walton, G.N. *A computer algorithm for Predicting infiltration and inter-room airflows.* ASHRAE Transactions, 1984, Vol. 90, Part. 1B, p. 601-609.

32. Roldan, A. *Etude thermique et aéraulique des enveloppes de bâtiment. Influence des couplages intérieurs et du multizonage.* Thèse Sci. Institut National des Sciences Appliquées de Lyon, 1985. 310 p.

33. Y. Li, A.E. Delsante, J.G. Symons, Simulation tools for analysing natural ventilation of large enclosures with large opennings. *AIRAH Journal*, pp. 21-28, November 1997.

34. R. Fauconnier, P. Guillemard, A. Grelat, Algorithmes des simulateurs du comportement thermique des bâtiments BILGA et BILBO, Annales de ITBTP, n° 458, Oct. 1987 (1987)

35. Feustel, H., E., Dieris, J. *A survey of airflow models for multizone structures.* Energy and Buildings, 1992, n°18. p. 79-100.

36. J. Verbeke, R. Cools, The Newton-Raphson method, *Int. Jour. of Mathematical Education*, 26, 2, 1995

37. G.N. Walton, A computer algorithm for predicting infiltration and interroom airflows. *ASHRAE Transactions* 90, Part 1, 601-610 (1984)

38. Herrlin, M., K., Allard, F. *Solution methods for the air balance in multizone buildings.* Energy and Buildings, 1992, n°18. p. 159-170.

39. H. Boyer, F. Garde, J.C. Gatina J. Brau, A multi model approach of thermal building simulation for design and research approach. Energy and Building, 28, (1998), pp. 71-78.

40. H. Boyer, J.C. Gatina, J. Brau, F. Pignolet, Modelisation methods and data structuration induced for simulation codes, *In Proceedings of First Joint Conference of International Simulation Societies*, ETH, Zurich, pp.729-733, 22-25 Août 1994

41. W.H. Press, Flannery B., B.P. Teukolsky, W.T. Vetterling, Numerical recipes in C. The art of scientific computing, Cambridge University Press.

42. F Garde, Validation et développement d'un modèle thermo-aéraulique de bâtiments en climatisation passive et active. Intégration multimodèle des systèmes. *PhD Thesis*, Université de la Réunion, France (1997)

43. F. Lucas, T. Mara, F. Garde, H. Boyer, A comparaison between *CODYRUN* and *TRNSYS*, simulation codes for thermal building behaviour, To appear in Renewable Energy

44. H. Boyer, Conception thermo-aéraulique de bâtiments multizones. Proposition d'un outil à choix multiple des modèles. *PhD Thesis*, Institut National des Sciences Appliquées de Lyon, France (1993)

45. F. Garde, H. Boyer, J.C. Gatina, Demand side management in tropical island buildings. Elaboration of global quality standards for natural and low energy cooling in buildings, Accepted, *Building and Environment*

**FIGURES**

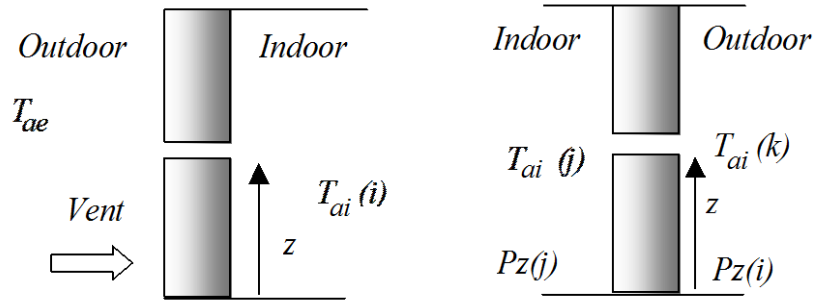

Fig 1 : *Exterior and interior small openings*

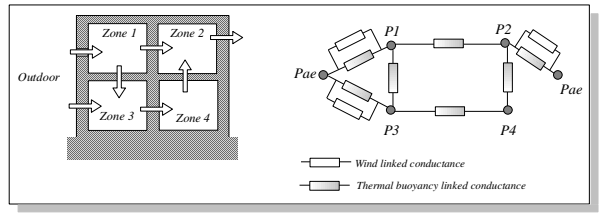

Fig 2 : *Airflow network*

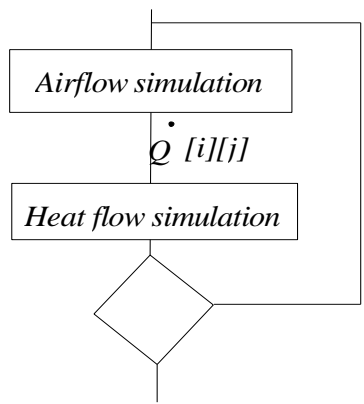

Fig. 3 : *Coupling flow chart*

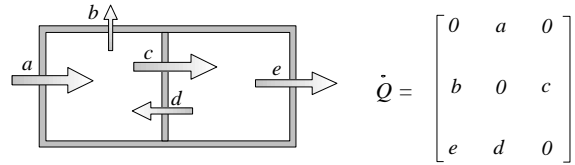

Fig. 4 : *Interzonal airflow rates matrix*

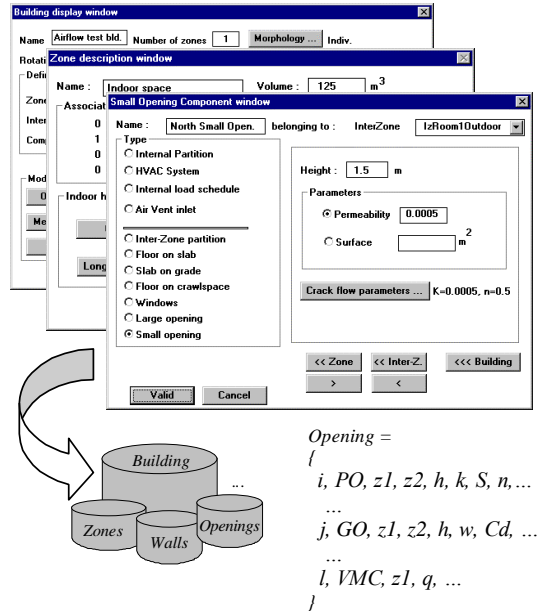

Fig. 5 : *The description windows, files and Openings data strucrure.*

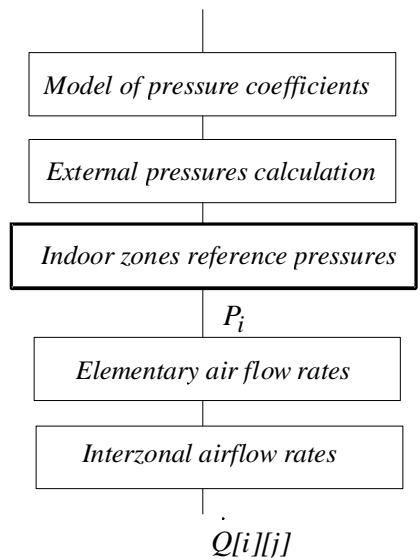

Fig. 6 : *Flow chart for airflow rates calculation*

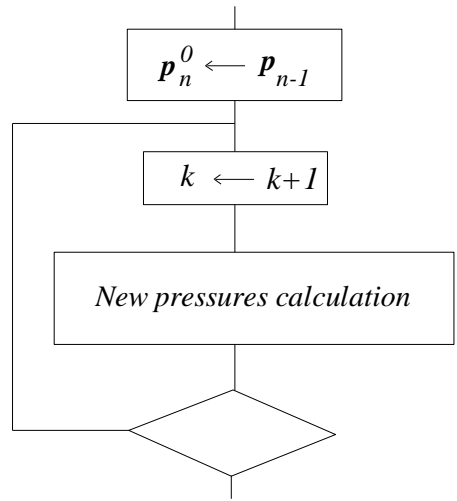

Fig. 7 : *Non linear pressure system solution*

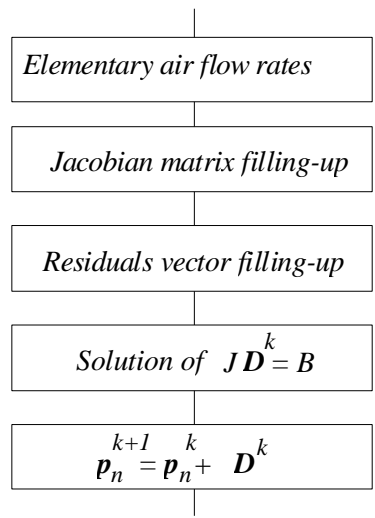

Fig. 8 : *Pressure calculation flow chart*

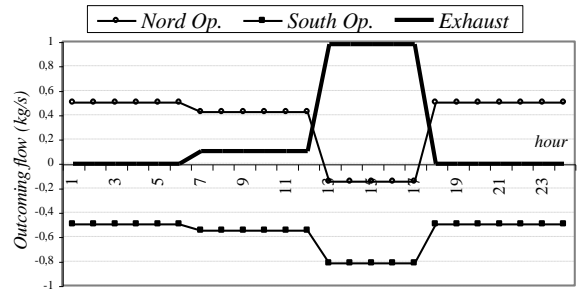

Fig. 9 : *Evolution of outcoming airflows*

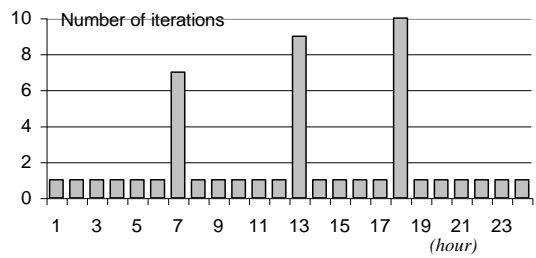

Fig. 10 : *Required iterations*

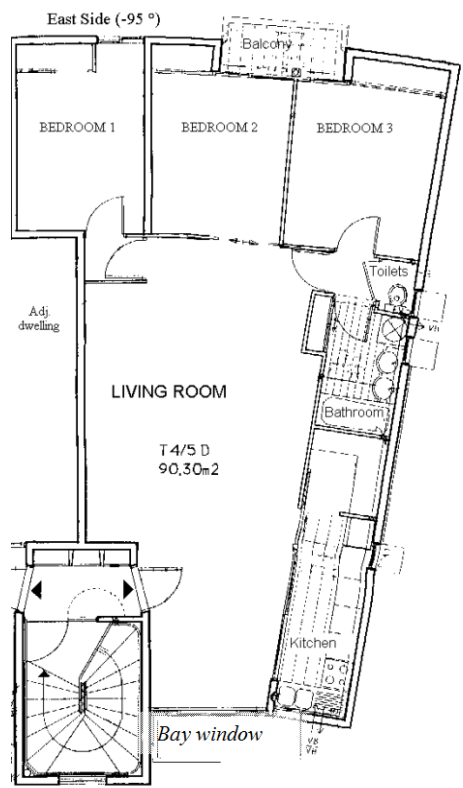

Fig. 11: *Typical dwelling*

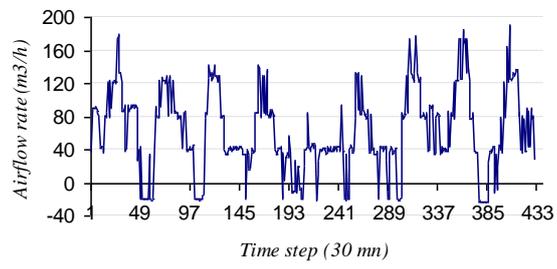

Fig. 12 : *Incoming air flow in the living room*

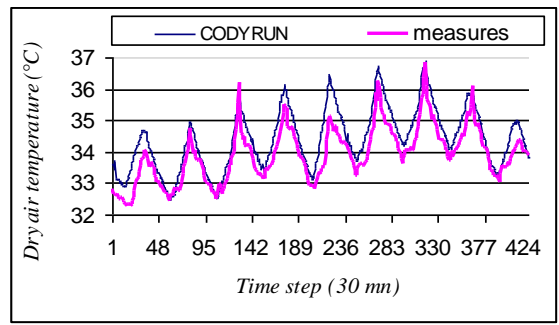

Fig .13 : *Dry air temperature drawings*